\newcommand{\beq}{\begin{equation}}
\newcommand{\eeq}{\end{equation}}
\newcommand{\beqns}{\begin{equation}}
\newcommand{\eeqns}{\end{equation}}
\newcommand{\beqar}{\begin{eqnarray}}
\newcommand{\bs}{\begin{eqnarray*}}
\newcommand{\eeqar}{\end{eqnarray}}
\newcommand{\es}{\end{eqnarray*}}
\newcommand{\beqml}{\begin{mathletters}}
\newcommand{\eeqml}{\end{mathletters}}
\newcommand{\phihat}{\hat{\varphi}}
\newcommand{\etahat}{\hat{\eta}}
\newcommand{\zetahat}{\hat{\zeta}}

\newfont{\fancy}{msbm10 scaled\magstep1}
\documentstyle[preprint,aps]{revtex}
\begin{document}
\draft
\title{Real Space Renormalization Group for
Langevin Dynamics in Absence of Translational Invariance}
\author{Achille Giacometti$^{(1)}$, Amos Maritan$^{(2)}$, Flavio
Toigo$^{(3)}$, Jayanth R. Banavar$^{(4)}$}
\address{$^{(1)}$Dipartimento di Fisica dell'Universit\'a di Padova, via
Marzolo 8, 35100 Padova, Italy \\
Present address: Institut f\"{u}r Festk\"{o}rperforschung
der Kernforschungsanlange, \\
Postfach 1913, D-52425, J\"{u}lich, Germany}
\address{$^{(2)}$Dipartimento di Fisica dell'Universit\`a  and
Istituto Nazionale di Fisica Nucleare, \\
Sezione di Padova, via Marzolo 8, 35100 Padova, Italy}
\address{$^{(3)}$Dipartimento di Fisica dell'Universit\`a  and \\
 Istituto Nazionale di Fisica della Materia INFM,\\
via Marzolo 8, 35100 Padova, Italy}
\address{$^{(4)}$Department of Physics and Center for Materials Physics,
104 Davey Laboratory,\\
The Pennsylvania State University,
University Park, PA 16802}
\date{\today}
\maketitle
\begin{abstract}
A novel exact dynamical real space renormalization group for a Langevin
equation
derivable from a Euclidean Gaussian action is presented.
It is demonstrated rigorously that an algebraic temporal law holds for
the Green function on arbitrary structures of infinite extent.
In the case of fractals it is shown on
specific examples that two different fixed points are found at variance
with periodic structures. Connection with growth dynamics of
interfaces is also discussed.
\end{abstract}
\narrowtext
\section{Introduction}
\label{sec:introduction}
Dynamical Renormalization Group (DRG) analyses of the critical dynamics of
statistical mechanics models has a long history \cite{HOH,MAZ,NG} with
virtually all studies on regular translationally invariant structures.
These approaches were, both in real and in momentum
space, of a perturbative nature; in contrast, the principal aim of the
present paper is the implementation of an {\em exact} approach, in real space,
for the dynamics occurring on fractal structures.

Consider a system described by a
Hamiltonian (or action) $H(\{ \varphi \})$
with the field variables $\varphi_x$ defined on the lattice sites $x$.
The simplest Langevin \cite{HOH} dynamics leading to equilibrium
with the correct Boltzmann weight $\exp [-H(\{\varphi\})]$ is \cite{Kardar}:
\begin{equation}
{\partial \varphi_x (t)\over\partial t}  = - {\delta H(\{\varphi\})
\over \delta \varphi_x} + \eta_x(t)\ .
\label{langevin}
\end{equation}
The temperature has been absorbed in the definition of $H$. The stochastic
noise $\eta_x(t)$ is chosen from a Gaussian distribution
\beqar
{\cal P}(\{ \eta(t) \}) &=& {\cal N}^{-1} \exp(-\sum_{x} \int \; d \tau
\frac{\eta_{x}^2(\tau)}{4 D})
\label{noise_dist}
\eeqar
(${\cal N}^{-1}$ is a normalization factor) which has a zero
average and variance $<\eta_x(t) \eta_y(t')> = 2 D \delta_{x,y} \delta(t-t')$.

We consider the simplest possible choice of the Hamiltonian
$H(\{\varphi\})$, the Gaussian model:
\begin{equation}
H(\{ \varphi\}) = \sum_x {a_x\over 2}\varphi^2_x -\sum_{<xy>}
\varphi_x\varphi_y,
\label{gaussian}
\end{equation}
where $a_x$ depends on temperature and the scale of $\varphi_x$ is
such that the second term in (\ref{gaussian}), i.e. the sum over nearest
neighbour sites, has coefficient 1.

Besides being the starting point
of more complicated models, the Gaussian model is related to many physical
situation such as the properties of an ideal polymer in solution \cite{deG},
diffusion processes and the growth dynamics of interfaces \cite{EW}.
Indeed if $a_x=k^{-1}$ for any $x$, then the hamiltonian (\ref{gaussian})
describes equilibrium properties of an ideal linear polymer in
solution with a monomer fugacity equal to $k$. In $a_x=z_x$, the coordination
number of site $x$, the hamiltonian (\ref{gaussian}) can be rewritten
as :
\beqar \label{gaussian2}
H&=& \frac{1}{2} \sum_{<x,y>} (\varphi_x-\varphi_y)^2
\eeqar
and it is related to the diffusion process known as the
{\it ant in the labyrinth} \cite{deG,Kehr}.

If $\varphi_x$ represents the height of an interface above the substrate
point $x$ (in the solid-on-solid approximation), then eq. (\ref{gaussian})
can be interpreted as the energy of an interface and the Langevin
eq. (\ref{langevin}) describes its dynamics. In a regular (hyper)cubic
lattice this is known as the Edward-Wilkinson model for interface
dynamics \cite{EW}.

We shall show, in specific examples,
that in the presence of non-translational structures (such as fractals)
even the simple model (\ref{gaussian}) gives rise to interesting behaviours.
Specifically
we will show that {\em two} different fixed points are present
unlike in structures with translational invariance where they collapse
to the same fixed point. Part of the results presented here has already
appeared in reference \cite{MTB}.

The paper is organized as follows. In Sec.\ref{sec:onedimensional}
we introduce the DRG technique for this problem first in the simple case
of a one-dimensional lattice and then in a situation in which the couplings
have a hierarchical structure. In Sec. \ref{sec:fractals}
the non-trivial cases of fractal structures with uniform and non-uniform
coordination number are analyzed. In the latter case it will be shown
that there is an additional universality class.  Sec.
\ref{sec:generalization} contains both rigorous and heuristic arguments
on a general network and for a non-linear case. The heuristic
arguments are then checked by the numerical analysis of Sec.
\ref{sec:numerical}. Finally in Sec.\ref{sec:conclusions} some
closing conclusions are presented.

\section{One-dimensional lattice}
\label{sec:onedimensional}
We shall start with the simple case of a one-dimensional lattice.
After a pedagogical example with uniform couplings, we
will turn to a non-trivial example. In the latter case only
DRG allows to obtain the asymptotic solution.
\subsection{Uniform couplings}
\label{subsec:uniform}
Let us start with a one-dimensional case in order to show how
the method works. Equation (\ref{langevin}) for the Hamiltonian
(\ref{gaussian}) with $a_x=a$ has then the form:
\beqar \label{eq_1d}
\frac{\partial \varphi_x(t)}{\partial t} &=& (\varphi_{x-1}(t)+
\varphi_{x+1}(t)-a \varphi_x(t)) + \eta_x(t)
\eeqar
It is immediately clear that if we were to choose uncorrelated noise,
decimation would produce correlation between
the nearest-neighbours noises. We thus assume a nearest-neighbour
correlation to start with, that is:
\beqar \label{noise_corr2}
<\eta_x(t_1) \eta_y(t_2)> = 2 D_{x,y} \delta(t_1-t_2)
\eeqar
where $D_{x,y}=D_0 \delta_{x,y}+D_1 \delta_{x \pm 1,y}$.
As usual it is convenient to work in (time) Fourier space:
\beqar \label{momentum_1d}
\alpha \phihat_x(\omega) &=& \phihat_{x-1}(\omega)+
\phihat_{x+1}(\omega) + \etahat_x(\omega)
\eeqar
where:
\beqar \label{Fourier}
f_x(t) &\equiv& \int_{-\infty}^{+\infty}\;\; \frac{d \omega}{2 \pi}
e^{-i \omega t} \hat{f}_x(\omega)
\eeqar
and $\alpha=a-i \omega$.

Upon decimation of (e.g.) the odd sites in favour of the even ones
(i.e. solving eq. (\ref{momentum_1d}) for $\phihat_{2x\pm1}$ in
terms of $\phihat_{2x}$ and $\phihat_{2x \pm 2}$),
we get an equation for the surviving (even) sites:
\beqar \label{eq_even}
(\alpha^2-2) \frac{\phihat_{2x}(\omega)}{\alpha} &=&
\frac{(\phihat_{2x-2}(\omega)+\phihat_{2x+2}(\omega))}{\alpha}+
\zetahat_{2x}(\omega)
\eeqar
where we have defined a new noise $\zetahat_{x}(\omega)$ as
\beqar \label{new_noise}
\zetahat_{2x}(\omega)&=& \etahat_{2x}(\omega) +
\frac{(\etahat_{2x-1}(\omega)+\etahat_{2x+1}(\omega))}{\alpha}
\eeqar
which is correlated as
\beqar \label{corr_new_noise}
<\zetahat_{x}(\omega_1)\;\;\zetahat_{y}(\omega_2)>&=& 2 \tilde{D}_{x,y}
(2 \pi) \delta(\omega_1+\omega_2)
\eeqar
where as before $\tilde{D}_{x,y} =\tilde{D}_0 \delta_{x,y}+
\tilde{D}_1 \delta_{x \pm 1,y}$ and
\beqml
\beqar
\tilde{D}_0 &=& D_0 (1+\frac{2}{\alpha^2})+D_1 \frac{4}{\alpha}
\label{tilde:1} \\
\tilde{D}_1 &=& D_0 \frac{1}{\alpha^2}+
D_1 \frac{2}{\alpha} \label{tilde:2}
\eeqar
\eeqml
Therefore the new noise keeps the same correlation as the original
one and being a linear combination of Gaussian noises is itself a
Gaussian noise. With the redefinitions:
\beqml
\beqar
\phihat_{2x}(\alpha,D) &=& \alpha A \phihat_{x}(\alpha',D')
\label{ren:1} \\
\alpha' &=& \alpha^2-2 \label{ren:2} \\
\zetahat_{2x} &=& A \etahat_{x'}^{\prime}
\label{ren:3}
\eeqar
\eeqml
the Langevin equation preserves its original form.
The static recursions relation for $a$ is obtainable from (\ref{ren:2})
for $\omega=0$, i.e.
\beqar \label{static}
a^{\prime} &=& a^2 -2
\eeqar
from which the recursions equation for $\omega$ is derivable. Indeed
if $\alpha=a-i \omega$, then $\alpha^{\prime}= a' -i \omega^{\prime}$
with:
\beqar \label{rec_omega}
\omega^{\prime} &=& 2a\omega-i\omega^2
\eeqar

The variance of the new noise is
$<\etahat_{x}(\omega_1')\;\etahat_{y}(\omega_2')> = 2 D_{x,y}' (2 \pi)
\delta(\omega_1'+\omega_2')$ with $D'(\omega')=  T(\omega)
\cdot D(\omega)$.
Here we have defined the $1 \times 2$ matrix $ D = \pmatrix{&D_{0}
\cr &D_{1}\cr}$
and $ T$ is the transmission matrix which, in the long time limit
$\omega \rightarrow 0$ becomes:
\beqar \label{transmission}
 T &=& \frac{2a}{A^2}
\: \left( \begin{array}{cc}
1+\frac{2}{a^2} & \frac{4}{a} \\
\frac{1}{a^2} & \frac{2}{a}
\end{array} \;\; \right)
\eeqar
It is important to note that the amplitude $A$, which is always allowed
since it drops out from the
final equations, is determined by the fixed point condition on $D$.
The critical fixed point of equation (\ref{static}) is $a^{*}=2$
(model (\ref{gaussian}) in 1-d is meaningful only for $a \geq 2$).
Since we are interested in critical dynamics, we will put $a=a^*=2$
in what follows.
By setting the determinant of
the system giving the fixed point for the matrix $D$ equal to zero one
obtains two solutions (two lines of fixed points corresponding
to fixed ratios $D_0/D_1$) associated
respectively to $A^{*}=2^{1/2}$ and $A^{*}=2^{3/2}$. It is easy
to see that the first choice leads to an unstable solution while
the second results in a stable one. Indeed, let $\Lambda_{+}$
and $\Lambda_{-}$ be the
two eigenvalues of the transmission matrix $ T$ and $u_{+}$ and
$ u_{-}$ the corresponding eigenvectors. Then if we start
with an intial state $D$ :
\beqar \label{expansion}
 D &=& c_{+}  u_{+} + c_{-}  u_{-}
\eeqar
($c_{\pm}$ being the coefficients of the expansion in the basis
of eigenvectors) upon iteration of the RG transformation,
the variance grows unbounded with the choice $A^{*}=2^{1/2}$
while it is driven toward a stable fixed point $ D^{*}= c_{+} u_{+}$
with the choice $A^{*}=2^{3/2}$.

We now have the means to compute the critical exponents of interest.
{}From eq.\ (\ref{ren:1}) we have, at the stable fixed point, for a generic
scaling factor $b>1$:
\beqar \label{scaling_field}
\phihat_{bx}(\omega,D^{*}) &=& b^{\phi}
\phihat_x(b^{z} \omega,D^{*})
\eeqar
where $\phi=5/2$,$z=2$ is the dynamical exponent $z$ and we have written
$\phihat(\omega,D)$ instead of $\phihat(2-i\omega,D)$.
The exponent $\phi$
can be related to the scaling of
the two-point correlation function:
\beqar \label{correlation_function}
G_{x,y}(t_1,t_2) &=& <\varphi_x(t_1)\; \varphi_y(t_1)>
\eeqar
where the average is over noise configurations.
Using eq.(\ref{scaling_field}), we get:
\beqar \label{scaling_corr}
G_{bx,0}(t_1,t_2)&=& b^{2(\phi-z)} G_{x,0}(\frac{t_1}{b^z},\frac{t_2}{b^z})
\eeqar
The properties of the {\em equal time} correlation function can now
be easily derived. If we define the function $W^2(L,t) \equiv G_{L,0}(t,t)$
we end up with the following scaling form:
\beqar \label{scaling_width}
W(L,t) &=& L^{\alpha} F(\frac{t}{L^z})
\eeqar
with $\alpha=\phi-z=1/2$ and $z=2$.

Few comments are in order. If $\varphi_x$ represents the height
of an interface above the substrate site $x$, eq.(\ref{eq_1d}),
with $a=a^*=2$
coincides with the growth equation proposed by Edwards and Wilkinson
\cite{EW} with unit surface tension $\nu$, namely:
\begin{equation} \label{eq_EW}
{\partial \varphi_x(t)\over\partial t} = \nu \nabla^2 \varphi_x(t)
+\eta_x(t)
\end{equation}
where $\nabla^2$ is the discrete laplacian defined as
\beqar
\nabla^2 \varphi_x &=& \sum_{y(x)} (\varphi_y-\varphi_x)
\eeqar
Here notation $y(x)$ means that the sum is restricted to the
nearest neighbours of $x$.
Then $W(L,t)$ coincides
with the width of the interface which, as a function
of time and of the lateral extent of the substrate $L$, is found to
exhibit  the general scaling form \cite{vicsek}
$W(L,t)\propto L^{\alpha} f(t/L^z)$,  which reduces
to $W(L,t) \propto t^\beta$ for $t<<L^z$ and to $W(L,t)\propto L^\alpha$
for $t>>L^z$, where $\alpha=\beta z$ and $z$ is the standard dynamical exponent
\cite{HOH}.

For a $d$-dimensional substrate Edwards and Wilkinson \cite{EW}
found $\beta=\max \left({2-d\over 4}, 0\right)$ and $z=2$ with an
upper critical dimension of 2. Above $d=2$ the interface is almost
flat with $\alpha=\beta=0$.
Our results on a one-dimensional case, thus coincide with the exact results.
We shall now apply the same methodology to
a non-trivial case.
\subsection{Hierarchical couplings}
\label{subsec:hierarchical}
A rather interesting case which lacks translational invariance
and which can be solved only by RG techniques,
is the case of a one-dimensional lattice where however the coupling
between nearest-neighbours are hierarchically distributed as
in Figure \ref{fig1}. In the case of diffusion, those kind
of models have already been studied \cite{HK,Giacometti}.

In the present context we start from the modified Hamiltonian:
\begin{equation}
H(\{ \varphi\}) = \sum_x {a_x\over 2}\varphi^2_x -\sum_{<xy>} w_{x,y}
\varphi_x\varphi_y
\label{gaussian_2}
\end{equation}
where $w_{x,y}=w_{y,x}$ have the hierarchical
structure defined as follows (see Fig.\ref{fig1}):
\beqar \label{fugacities_hierarchical}
 w_{x,x-1}=w_{x-1,x} &=& \left\{ \begin{array}{ll}
                \epsilon_n \; \; & \mbox{$n \geq 1$} \\
                \epsilon_0=1
                \end{array}
                \right.
\eeqar
where $x=(2m+1)2^n \;\;\; (n,m=0,1,2,...)$ and $w_{x,y}=0$
if $|x-y|>1$. The choice $\epsilon_0=1$ can be made without loss of
generality by a proper rescaling of the time scale.

As in the previous example we will work at criticality. This
corresponds to the choice $a_x=\sum_{y=x \pm 1} w_{x,y}$,
i.e.
\beqar
H &=& \frac{1}{2} \sum_{<x,y>} w_{x,y} (\varphi_x-\varphi_y)^2
\eeqar

The implementation of the RG transformation closely
follows the one used in the diffusion case \cite{HK,Giacometti}
and it is based on the decimation of the sites indicated by
a circle in Fig.\ref{fig1}. Interestingly, due to the particular
decimation scheme chosen, the minimal set of parameters for the variance
which are invariant under RG transformation, is $D_{x,y}=D_0 \delta_{x,y}
+D_{1,2}\delta_{y,x \pm 1}$ where one uses $D_1\;(D_2)$ if $w_{x,y}=\epsilon_0
\;(\epsilon_n)$ respectively. In other words a single self-correlation
but two different nearest-neighbour correlations are involved.
The recursions turn out to be:
\begin{mathletters}
\begin{equation}
\phihat_{bx} (\omega,  D) = A {(\alpha_1^2 -\epsilon_1^2)
\over\epsilon_1}  \phihat_x (\omega',D')
\label{rg_HK:a}
\end{equation}
\begin{equation}
\epsilon_n' = \epsilon_{n+1} \frac{(\alpha_1^2-\epsilon_1^2)}{\epsilon_1}
\;\; (n=1,2,....)
\label{rg_HK:b}
\end{equation}
\begin{equation}
\omega' = \frac{2(1+2 \epsilon_1)}{\epsilon_1} \omega + 0(\omega^2)
\label{rg_HK:c}
\end{equation}
\begin{equation}
A\etahat_{x'} (\omega') =\etahat_{x'} (\omega) +
\frac{\etahat_{y_L} (\omega) \epsilon_1 + \etahat_{y_R} (\omega) \alpha_1}{
(\alpha_1^2-\epsilon_1^2)}
\label{rg_HK:d}
\end{equation}
\end{mathletters}

\noindent
where $\alpha_1=-i\omega+1+\epsilon_1$ and $y_L(y_R)$ are
left (right) sites which are decimated with respect to the
barrier $\epsilon_1$.
Apart from the noise contribution, the recursions are the
same as in the case of diffusion \cite{HK,Giacometti}.
It proves convenient to define $R_n=\epsilon_{n+1}/\epsilon_n$
since, upon iteration of eq.(\ref{rg_HK:a}), it flows to a fixed
point $R \in (0,1]$ and then the analysis of the
fixed points is reduced to a study of a two-dimensional map in the
$\{\epsilon_1,\omega \}$ plane. There are two fixed points
O$=(\epsilon_1^{*}=+\infty, \omega^{*}=0)$ and
A$=(\epsilon_1^{*}=R/(1-2R),\omega^{*}=0)$ whose stability depends upon
the value of $R$. We will distinguish two cases:
\begin{description}
\item[A) $R<1/2$]: A is stable and O is unstable and
$y=d_W=\ln (2/R)/\ln 2$. The $\omega \rightarrow 0^{+}$ limit
of the transition matrix $T$ is
$T=(2/R)A^{-2} \pmatrix{&a(R) &b(R) &c(R) \cr
                          &c(R) &d(R) &e(R) \cr
                          &0 &0 &1 \cr}$
where $a(R)=2(1-R+R^2)$, $b(R)=2(1-R)$, $c(R)=2R(1-R)$, $d(R)=2R$,
$e(R)=1-2R+2R^2$. The value of $A^{*}$ which yields a stable
fixed line is $A^{*}=(4/R)^{1/2}$. Then, proceeding as in the
previous example, one finds again $\alpha=\frac{y-1}{2}=
\frac{|\ln R|}{2\ln 2}$ as expected since in this case
$d_f=d=1$.
\item[B) $R>1/2$]: O is stable and A is unstable and then $y=d_W=2$
as in the equal coupling case. The matrix $T$ is
$T=4A^{-2} \pmatrix{&\frac{3}{2} &1 &\frac{1}{2} \cr
                          &\frac{1}{2} &1 &\frac{1}{2} \cr
                          &0 &0 &1 \cr}$
which yields $A^{*}=2^{3/2}$ (again as in the equal coupling case)
and consequently $\alpha=1/2$.
\end{description}

\section{Fractals}
\label{sec:fractals}
An interesting non-translationally invariant case, which can be
analyzed with the technique described above, is the case of fractal
structures. We will consider several examples which are prototypes of
different physical situations.
\subsection{Fractals with uniform coordination numbers}
\label{subsec:sierpinski}
The first example is the standard Sierpinski Gasket shown in Fig.\ref{fig2}
whose fractal dimension is $d_f=\ln 3/\ln 2$. In this case, as on
translationally invariant structures, there is only one fixed point.
The renormalization group procedure is readily implemented in the
Fourier transformed equation of motion (\ref{langevin}) for the
hamiltonian (\ref{gaussian}) which reads:
\beqar \label{eq_SG}
\alpha_x \phihat_x(\omega) &=& \sum_{y(x)}\phihat_{y}(\omega)+
 \etahat_x(\omega)
\eeqar
where $\alpha_x=a_x-i \omega$ and the sum is over nearest neighbours
of $x$. The set up for the noise is the same as in the one-dimensional case.
As before we put directly $a=a^*=4$ corresponding to the static
fixed point.
Then, following standard procedure,
the RG transformation is carried out in two steps:
\begin{description}
\item[a)] The $\phihat_x$ on the circled sites in Fig.\ \ref{fig2} are
eliminated in order to rescale the system by a scale factor $b$
($=2$ in this case).
This may be done by solving Eq.\ (\ref{eq_SG}) for $\phihat_x (x=1,2,3)$
in terms of $\phihat_y$ $(y=A,B,C)$ (see Fig.\ \ref{fig2}) and
substituting the resulting equations back into (\ref{eq_SG}).
\item[b)] $\phihat_x$ and $\etahat_x$ are then suitably rescaled as follows:
\begin{mathletters}
\begin{equation}
\phihat_{bx} (\omega, D) = A (\omega+6)
\phihat_x (\omega',D')
\label{rg_SG:a}
\end{equation}
\begin{equation}
A\etahat_{x'} (\omega') =(\omega+5)(\omega+2)\etahat_{x'}(\omega) +
2 \sum_{y(x')} \etahat_y (\omega) +
(\omega+4) \sum_{y((x'))}\etahat_{y} (\omega)
\label{rg_SG:b}
\end{equation}
and
\begin{eqnarray}
\omega' &=& \omega(5-i\omega)
\label{rg_SG:c}
\end{eqnarray}
\end{mathletters}

\noindent
where primed (unprimed) sites refer to the surviving (decimated) sites under
the RG transformation, and $y(x)$ and $y((x))$ are the decimated
nearest-neighbours and the decimated next-nearest-neighbours respectively.
\end{description}
The procedure for calculating the amplitude $A^{*}$ corresponding
to the stable fixed point proceeds along exactly the same lines
as the one-dimensional case: again the minimum set of
parameters which are invariant under the RG transformation for the noise
is $\{D_0,D_1 \}$ corresponding to the self and nearest-neighbour
correlation respectively. It is noteworthy that the presence of
noise does {\em not} influence the renormalization of
time, thus prompting the identification of the
dynamical critical exponent $z$ with the fractal dimension $d_W$
of the walk on that structure giving the end-to-end distance $R(t)$
of the {\it ant in the labirinth} at large $t$, i.e. $R(t) \sim t^{1/d_w}$.
This is given by (\ref{rg_SG:c}) in the $\omega \rightarrow 0$ limit
as $\omega^{\prime}= b^{d_w} \omega$.
One then finds the following critical exponents:
\begin{equation}
z = d_W \quad , \quad \beta=\frac{2-d_S}{4} \quad , \quad
\alpha=\frac{2-d_S}{4}d_W =\frac{d_W-d_f}{2}
\label{esp_SG}
\end{equation}
with $d_W=\ln 5/\ln 2$ and $d_S=2 d_F/d_W=\ln 9/\ln 5$ is the spectral
dimension
describing the low frequency behaviour of the density of vibrational
modes \cite{AO,Rammal}.
A similar analysis can be carried out for a general d-dimensional
Sierpinski Gasket.
\subsection{Fractals with non-uniform coordination number}
\label{subsec:T-fractals}
In the previous example we investigated the effect of the
self-similarity of the structure on the scaling of the two-point
correlation function for a Langevin equation induced by a Gaussian
Hamiltonian. However in that example the coordination number
was uniform (equal to $4$) as in the translationally invariant lattice.
The interplay between  self-similarity of the substrate and
non-uniformity of the coordination number has been recently investigated
\cite{GIA}.
Contrary to what happens on periodic structures, it has been shown
that on both statistical and deterministic
fractal structures the equilibrium properties of (\ref{gaussian}) are
governed by two universality classes, corresponding to ideal
polymers and to the ant in the labyrinth \cite{GIA}.
We now study this effect on the Langevin equation on specific
examples. The T-fractal (Fig.\ref{fig3}) has $z_x-1,3$ whereas
the Branching-Koch-Curve (BKC) (fig. \ref{fig4}) has $z_x=2,3$.

We stress again that the appearance of two fixed points is a feature of
non-uniform coordination number and a fractal structure.

In the recursion relations we will derive below, the (minimum) set
of parameters invariant under RG transformation is
given by $D_{x,y}=D_{0z}$ if $x=y$ ($z$ is the coordination number), and
$D_{x,y} =D_1$ if $x$ and $y$ are nearest neighbors and
$D_{x,y}=0$ otherwise. We will write $D =
\pmatrix{&D_{01}\cr &D_{03}\cr &D_1\cr}$ and $\alpha =
\pmatrix{&\alpha_1 \cr &\alpha_3\cr}$.
With this distinction in mind, calculations along the same lines
as before, yield the following recursions:
\begin{mathletters}
\label{rg}
\begin{equation}
\phihat_{bx} (\alpha, D) = A {\alpha_1 \alpha_3 - 1
\over\alpha_1} \ \phihat_x (\alpha',D')
\label{rg:a}
\end{equation}
\begin{equation}
A\etahat_{x'}^{\prime} (\alpha') =\etahat_{x'} (\alpha) +
\sum_{y(x)} \big(\alpha_1
\etahat_{y_1} (\alpha) + \etahat_{y_3} (\alpha)\big) \big/
(\alpha_1\alpha_3 -1)
\label{rg:b}
\end{equation}
and
\begin{eqnarray}
\alpha'_1 &&= \alpha_1\alpha_3 - 2 \nonumber\\
\alpha'_3 &&= \alpha^2_3 -\alpha_3/\alpha_1 - 3
\label{rg:c}
\end{eqnarray}
\end{mathletters}
In the above definitions $\etahat_{y_1,y_3}$
represents the noise associated with sites of coordination
$1$ and $3$ respectively.

In the static limit, $\omega = 0$, one has $\alpha_x=a_x$
and as found in \cite {GIA}, the recursion equation (\ref{rg:c})
admits two fixed points (see Fig \ref{fig5}): $G=(a_1^\ast, a_3^\ast)=(1,3)$,
corresponding to the ``ant in the labyrinth", and  $P=(a_1^\ast,
a_3^\ast)= \big(\infty, (1+\sqrt{13})/2\big)$, describing the scaling
behaviour of the ideal polymer \cite{GIA}.   (The symbols G and P denote
growth and polymer respectively.)  Indeed the linearized
recursions (it is better to work with $a_3$ and $a_3/a_1$) show that
$G$  is unstable (two relevant eigendirections) with a thermal exponent
$y_G=\ln 6/\ln 2$, which coincides with $d_W$ \cite{AO,Rammal}, the
random walk dimension, while $P$ is stable (one relevant and one irrelevant
eigendirection) with $y_P= {\ln (1 +\sqrt{13})\over \ln 2}$ related to
the end-to-end distance $R$ of an ideal polymer of length $N$ according to
$N\sim R^{y_P}$

    Now let us turn to the dynamics. In the long time limit, i.e.,
$\omega \to 0$, and near the above static fixed points one has $\alpha_x=
a^\ast_x{}$. This implies that under renormalization $\omega$
rescales as $\omega (b) \sim b^y \omega$: thus the dynamical
exponent $z$ is equal to the thermal exponent $y$ appropriate to the
fixed point. From eq. (\ref{rg:b}) one can easily calculate the
new noise correlation functions near the fixed point. The recursion
equation is:

\begin{equation}
 D' = T \cdot D
\label{td}
\end{equation}
with $T=6 A^{-2}  \pmatrix{ &{3\over 2} &{1\over 4} &{1\over 4} \cr
                         &{3\over 2} &{5\over 4} &{1\over 4} \cr
&{9\over 2} &{3\over 4} &{7\over 4} \cr}$
at the fixed point $G$ and
$T=2a^{-1}A^{-2} \pmatrix{&2a &0 &a^2 \cr
                          &2a &1 &a^2 \cr
                          &6a &0 &1+3a \cr}$
with $a^{-1}={1+\sqrt{13}\over 2}$ at the fixed point $C$.

   Thus, choosing the value of $A$ so that the the largest eigenvalue
is $1$, one finds that the corresponding eigenvector $D^\ast$
is a stable fixed point. Near the fixed points from eq. (\ref{rg:a})
one readily deduces
the scaling of $\phihat_x (\omega, D^\ast)$ (we are writing
$\phihat_x(\omega, D^\ast)$ instead of
$\phihat_x(a^\ast-i\omega, D^\ast)$) in the $\omega\to 0$
limit:
\begin{equation}
\phihat_{bx} (\omega, D^\ast) = b^{\alpha + z} \phihat_x (b^z
\omega, D^\ast)\ .
\label{sc}
\end{equation}

     From \ (\ref{rg:a}) and\ (\ref{sc}), one finds $(b=2)$
\begin{equation}
\alpha_G = {y_G-d_f\over 2}= \frac{1}{2} \quad \text{at}\ G\quad,\quad
\alpha_P={y_P-1\over 2}=0.601.. \quad \text{at}\ P
\label{esp}
\end{equation}
where $d_f=\ln 3/\ln 2$ is the fractal dimension of the structure.
We will show below that the expression for $\alpha_G$ is completely
general.
    Since $y_G=d_W$ and the spectral dimension, $d_S$, describing the
low frequency behaviour of the density of vibrational modes satisfies
the scaling relation \cite{Rammal} $d_S=2d_f/d_W$, the result\
(\ref{esp}) for $\alpha_G$ can be rewritten as $\alpha_G={2-d_S\over 4}
d_W$. Taking into account that $z=d_W$ one gets $\beta=(2-d_S)/4$.

An extension of the T-fractals to sites with coordination $z=1,4$
leads to exactly the same expressions as before
with the same numerical value of $\alpha_G=1/2$ with a
slightly higher value $\alpha_P=0.678...$ for the other fixed point.
Thus the $\alpha_G$ is the same as in 1-d model of Sec.
\ref{sec:onedimensional}.

A more interesting example because it is a structure with loops,
is the case of the Branching-Koch-Curve (see Fig.\ref{fig4}).
Calculations in this case are rather tedious but quite
analogous to the previous example and we will omit annoying details.
Also in this case it is sufficient to start with a variance
$D_{x,y}=D_{0z} \delta_{y,x} +D_1 \delta_{|x-y|,1}$, where $D_{0z}$
is defined as before and with  two $a$'s, $a_1$ and $a_2$ corresponding
to two types of coordinations.
At the fixed point G=$(a_2^*=2,a_3^*=3)$ one finds again
$\alpha_G=(y_G-d_f)/2$ with $y_G=d_W=\ln (40/3)/\ln 3$
and $d_f=\ln 5/\ln 3$ consistent with the previous claim
of a general form.
The other fixed point P$=(a_2^{*}=+\infty,a_3^{*}=\sqrt{5})$
yields $y_P=\ln 11/\ln 3$ and $\alpha_P=\ln 2/\ln 3=0.6309..$.
Notice that in the above example $\alpha_G \leq 1/2$ whereas
$\alpha_P >1/2$.
\section{Generalizations: general network and non-linearity}
\label{sec:generalization}
Eq.\ (\ref{langevin}) with $D_{xy} =D \delta_{xy}$ evolves to an equilibrium
state with a probability distribution for $\{ \varphi \}$ given by:
\begin{equation}
{\cal P}_{\text{eq}} (\{ \varphi\}) \propto \exp \left\{ {-\nu\over 2D}
\sum_{<xy>} (\varphi_x -\varphi_y)^2  \right\}
\label{peq}
\end{equation}
where $x$ and $y$ are nearest neighbor sites \cite{Gardiner}. The equilibrium
correlation function $<(\varphi_x-\varphi_0)^2>_T$ is defined as in eq.\
(\ref{correlation_function}) calculated with the weight \ (\ref{peq})
can be shown \cite{Giacometti} to be proportional to
the resistance $\Omega_{x,0}$ between two fixed sites $0$
and $x$ of the fractal network where conductances (equal to $\nu/D$)
connect nearest neighbor sites that is:
\beqar
<(\varphi_x(t)-\varphi_0(t))^2>_T &\sim& \Omega_{x,0} \sim |x|^{\zeta}
\eeqar
where the last equality defines the exponent $\zeta$ for the resistivity.
Scaling arguments predict that for fractals $\zeta= d_W-d_f$
(Einstein relation) \cite{gefen}.
Since the roughness exponent $\alpha$ is defined by
(see eq.(\ref{correlation_function},\ref{scaling_corr}) $<(\varphi_x
- \varphi_0)^2>_T \sim |x|^{2\alpha}$, one finds $\alpha = {d_W -d_f\over 2} =
{d_W(2-d_s)\over 4}$. On the basis of the previous exact RG, we argue
that the $\omega$ renormalization (Eqs.\ (\ref{rg}) and \ (\ref{sc})) is
not influenced by the noise term in Eq.\ (\ref{langevin}).
We also note that Eq.\ (\ref{langevin})  without the noise term is
merely the diffusion equation with $\varphi_x(t)$ interpreted as the
probability of finding a diffusing particle at $x$ at time $t$. If
$\omega$ renormalizes as $\omega'=b^z \omega$ to leading order, then the
mean square distance traveled after time $t$ behaves asymptotically
like $t^{2/z}$ \cite{Giacometti} implying $z=d_W$.

   We now show that the temporal behavior of the width of the
interface $W(t)$ for a substrate described by an arbitrary infinite
network is also given by $t^{(2-d_s)/4}$ (provided that $d_s$ can be defined
\cite{Cassi}).
This is a non-trivial result since we prove not only that the
temporal growth has a power law form, but also obtain the exponent.
The proof is heavily based on the rigorous results of Hattori et al.
(HHW) \cite{HHW}.
We assume as initial condition $\varphi_x (t=0)=0$. Let us go back
to $G_{xy}(t)=<\varphi_x(t) \varphi_y(t)>$ where the average is
over the noise as defined in
Sec.\ref{sec:introduction}. The formal (forward) solution of
equation (\ref{eq_EW}) can be written as:
\begin{equation}
\varphi_x (t) = \sum_y \int_0^{\infty} \; d\tau U_{xy} (t-\tau)\eta_y (\tau)
\label{hdit}
\end{equation}
where $U_{x,y}(t)$ is the retarded Green function, associated to (\ref{eq_EW}),
satisfying:
\beqar \label{retarded}
(\frac{\partial}{\partial t} -\nu \nabla_{x}^2)U_{x,y}(t) &=&
\delta_{x,y}\delta(t)
\eeqar
that is:
\beqar
U_{x,y}(t)&=& \theta(t) (e^{t \nu \nabla^2})_{x,y}
\eeqar
where $\theta(\cdot)$ is the step function.

Using (\ref{retarded}), we have:
\beqar \label{Green}
G_{x,x_0}(t) &=& <\varphi_x(t)\varphi_{x_0}(t)>=2 D \sum_y \int_{0}^{t}\;
d\tau \;\; U_{x,y}(t-\tau) U_{x_0,y}(t-\tau) \\ \nonumber
            &=& 2D \int_{0}^{\infty} U_{x,x_0}(2(t-\tau))
\eeqar
where we have used the condition $<\eta_x(t_1) \eta_y(t_2)>=2 D
\delta_{x,y} \delta(t_1-t_2)$.

Due to the initial condition $U_{x_0,x}(0^{+})=\delta_{x_0,x}$, then
$G_{x,x_0}(t)$ is a solution of:
\beqar \label{non-retarded}
(\frac{\partial}{\partial t} -4D \nu \nabla_{x}^2)G_{x,x_0}(t) &=& \delta_{x,y}
2D
\eeqar
It is then easy to verify that a probability $P_{x,x_0}(t)$ defined through
\beqar \label{probability}
G_{x,x_0}(t) &=& \int_{0}^{4Dt} \;d\tau P_{x,x_0}(\tau)
\eeqar
is a solution of the diffusion equation
\beqar
(\frac{\partial}{\partial t} - \nu \nabla_x^2) P_{x,x_0}(t) &=& 0
\eeqar
The results of Ref. \cite{HHW} can now be applied to Eq.\ (\ref{probability})
defined on the sites of an arbitrary graph. Since, under very general
conditions, $P_{x_0,x_0}(t) \sim t^{-d_s/2}$, from Eq.(\ref{probability})
we get:
\beqar \label{roughness}
W^2(L,t)|_{L=\infty} &=& G_{x_0,x_0}(t) \sim t^{2 \beta}
\eeqar
where $\beta=(2-d_s)/4$ as expected from the previously solved cases.

 From the same procedure one can recover also the exponent $\alpha$.
Indeed if we assume the standard {\it ansatz} which appears to be
valid for a generic fractal \cite{HBA}:
\beqar \label{ansatz}
P_{x,x_0}(t)&=& \frac{1}{t^{d_s/2}} \Phi(t|x-x_0|^{-y})
\eeqar
and using eq.(\ref{probability}) and (\ref{roughness}) one
finds $W(L,t)=L^{\alpha} f(t/L^z)$ with
$\alpha=d_W(2-d_S)/4$ as before (for $\alpha_G$ and $z=y$).

   We now turn to non-linear growth on a fractal. We conjecture that the
analog of the non-linear growth equation of reference \cite{KPZ} is given by:
\begin{eqnarray}
{\partial \varphi_x (t)\over \partial t} =&&
\nu \sum_{y(x)} [ \varphi_y(t) - \varphi_x(t)] \nonumber\\
&&+ \lambda \sum_{y(x)} [ \varphi_y(t) - \varphi_x(t)]^2 + \eta_x (t)\ .
\label{eqkpz}
\end{eqnarray}
where now $\varphi_x(t)$ is to be interpreted as the height of the
substrate at position $x$ at time $t$. The $\lambda$ term takes into
account the lateral growth of the aggregate \cite{KPZ}.
When $\lambda=0$ we recover Eq.\ (\ref{eq_EW}).  Under rescaling
of a factor $b$ of the length, the
left hand side and the first two terms of the right hand side scale as
$b^{\alpha-z}$, $b^{\alpha-d_W}$ and $b^{2 \alpha-d_W}$ respectively, where
$b^{-d_W}$ comes from the Laplacian on a fractal \cite{nota3}.
In the absence of the $\lambda$ term, dimensional analysis yields $z=d_W$,
(found above) whereas the presence of the $\lambda$ term leads
to the time derivative on the left hand side balancing the $\lambda$ term:
\begin{equation}
\alpha+z = d_W
\label{truesp}
\end{equation}
with the $\nu$ term being subdominant.
Eq.\ (\ref{truesp}) generalizes the exact equation \cite{meakin,forster}
for Euclidean lattices which are characterized by $d_W=2$.

    Additionally, following ref. \cite{Hentschel}, if we identify the
dimension of the noise term to be $b^{-(z+d_f)/2}$ (this is expected
to hold in the linear case where the $\lambda$ term is absent) and
extending the dimensional analysis, we find again eq.\ (\ref{esp})
for $\alpha_G$.

   In the presence of the $\lambda$ term, one may conjecture that the
noise term scales as $b^{-(z+\alpha d_c)/2}$ where $d_c$ is the chemical
dimension of the fractal \cite{Hentschel}. With this {\it ad-hoc} assumption
one finds for equation (\ref{eqkpz}):
\begin{equation}
z= d_W {2+d_c\over3+d_c},\qquad \beta = {1 \over 2+ d_c},\qquad
\alpha = {d_W\over 3+d_c}
\label{kpzesp},
\end{equation}
leading to:
$\alpha=0.5064$, $\beta = 0.2789$ in $d=2$, and $\alpha=0.5170$, $\beta
= 0.2500$ in $d=3$ Sierpinski gaskets. (For the gaskets in $d$
dimensions $d_c=d_f=\ln(d+1)/\ln2$ and $ d_W=\ln(d+3)/\ln2$ \cite{Rammal}).

   We stress the fact that Eq.\ (\ref{truesp}) and \ (\ref{kpzesp}) have
been derived simply on the basis of power counting, and reduce to
previously known approximations on translationally invariant structures
\cite{Hentschel,KK}.
\section{Numerical analysis}
\label{sec:numerical}
   In order to verify Eq.\ (\ref{truesp}) and check the conjecture
leading to Eq.\ (\ref{kpzesp}), we have carried out computer simulations of
growth models on Sierpinski gaskets.
We considered a sequence of sizes of gaskets in both two and three
dimensions, the largest one containing 1095 and 2050 sites respectively.
The value of $\alpha$ was estimated by comparing the scaling of the
saturated roughness with system size, whereas $\beta$ was deduced by
studying the temporal dependence of the roughness for the larger sizes.
(i.e. $W(L,t) \sim L^{\alpha}$ for $t>> L^z$ and $W(L,t) \sim t^{\beta}$
for $t<<L^z$).

To check the correctness of the programs, we first studied the linear
growth process in both $d=2$ and $3$ by adding and subtracting particles
on the gasket sites with the same probability and found excellent
agreement with the exact results of Eq.\ (\ref{esp}).

The nonlinear case  was then studied by carrying out simulations of the
Kim-Kosterlitz \cite{KK} growth model for the gaskets; the numerical results
confirmed the validity of Eq.\ (\ref{truesp}).
Specifically we found:
$\alpha=0.48 \pm 0.02$ and $\beta = 0.26 \pm 0.02$ in $d=2$, whereas
$\alpha=0.48 \pm 0.02$ and $\beta = 0.23 \pm 0.02$ in $d=3$.
\section{Discussions and Conclusions}
\label{sec:conclusions}
Our discussion so far has been restricted to growth on fractal substrates
with the particles arriving along a space dimension orthogonal to that
in which the fractal resides.
One might also consider growth on a self-affine substrate (with no overhangs,
so that all sites are accessible to the incoming particles) with the
growth occurring in the direction normal to the rough surface, but yet
in the $d$-dimensional hyperplane in which the surface resides.
In this case, the resulting growth would be characterized by the
regular exponents in $d-1$ dimensions.
This follows readily from two observations:

i) The exponent $\alpha$ characterizing the growth is a measure of the
equilibrium roughness and does not depend on whether the initial
configuration is a self affine or a planar interface.

ii) It is reasonable that the exponent $z$ which may be defined by studying
the relaxation of small perturbations around equilibrium is typically
the same that characterizes the approach to equilibrium from any initial
configuration.

The exponent $\beta$ is deduced directly from $\alpha$ and $z$.
Thus the growth exponents in this case are trivially determined.

Another observation is noteworthy. The results given here
are valid for systems of continuous symmetry spins where,
to the best of our knowledge,
no analog of Henley's argument, valid for
Ising spins undergoing Glauber dynamics, \cite{Henley} is known.

In conclusion we have presented a complete analysis of linear Langevin
dynamics.
The analysis has been carried out in real space and is based on
a combinations of RG analysis, rigorous results and heuristic considerations.

The primary results of our dynamical RG analysis are:

1. For eq.\ (\ref{eq_EW}), i.e. for linear \cite{EW} growth processes, the
width of the interface grows algebraically with time and the exponent
$\beta = {2-d_s\over 4}$, where $d_s$ is the spectral dimension of the
substrate. We stress that it is not common to explicitly demonstrate
in a rigorous way such a power law behavior.
Indeed, this result is valid not only for a fractal substrate, but for
any generic substrate: $d_s$ can be rigorously defined \cite{HHW}
for almost \cite{Cassi} all infinite graphs consisting of a set of nodes
and links joining sites that are defined to be nearest neighbors.

2.  For a fractal substrate, $z=d_W$, and $\alpha={2-d_s\over 4} d_W$,
where $d_W$ is the random walk dimension characterizing the asymptotic
behavior of the root mean square distance, $R$, traveled after a time
$t$, $R\sim t^{1/d_W}$.\\($d_W$, $d_s$ and the fractal dimension,
$d_f$, are related by $d_s=2 d_f/d_W$ \cite{DHAR,AO}).

3. For both fixed points of the hamiltonian (\ref{gaussian}) the dynamical
exponent $z$ coincides with the thermal critical exponent $y$ of the
corresponding static problem.

Result 1 is rigorous while 2 and 3 are based on exact arguments.
For non-linear growth model (\ref{eqkpz}), heuristic arguments
and numerical analysis suggests that eqs. (\ref{kpzesp},\ref{truesp})
are at least good approximations.

\section{acknowledgments}
   This work was supported by grants from NASA, NATO, NSF, ONR, the Donors
of the Petroleum Research Fund administered by the American Chemical
Society, The Center for Academic Computing at Penn State.


\begin{figure}
\caption{ Part of a one-dimensional lattice with hierarchical couplings
between the nearest-neighbours. Thicker bonds correspond to
weaker coupling. Circled sites are eliminated after one RG step which scales
the system by a factor of $2$.}
\label{fig1}
\end{figure}
\begin{figure}
\caption{ Part of an infinite Sierpinski Gasket in $d=2$. Circled sites
$1,2,3,...$ are eliminated in favour of surviving sites $A,B,C,..$
after one RG step which scales the system by a factor $2$.}
\label{fig2}
\end{figure}
\begin{figure}
\caption{ Part of an infinite T-fractal in $d=2$ which has coordination numbers
$z=1,3$. Circled sites are
eliminated after one RG step and this scales the system by
a factor $2$.}
\label{fig3}
\end{figure}
\begin{figure}
\caption{ Structure of the Branching-Koch-Curve in $d=2$ which has
$z=2,3$. Circled sites are eliminated after one RG step which scales
the system by a factor of $3$.}
\label{fig4}
\end{figure}
\begin{figure}
\caption{Flow diagram for the RG analisys for the T-fractal example.
Fixed points G and P correspond to {\it growth} and {\it polymer dynamic}
respectively. The line L is the critical line delimiting the
unphysical region.}
\label{fig5}
\end{figure}


\begin{references}
\bibitem{nota3} Since on an infinite lattice:
$\sum_{x} \varphi_x(t) \sum_{y(x)} [ \varphi_y(t) - \varphi_x(t)] =
 - \sum_{x} \sum_{y(x)} [ \varphi_y(t) - \varphi_x(t)]^2 $,
the scaling of the first two terms on the r.h.s. of Eq.\ (\ref{eqkpz}) differ
by a factor $b^{\alpha}$ (scaling of $\varphi$).
\end{references}
\end{document}